# Coupled cavity-waveguide system based on topological corner state and edge state

Aoqian Shi,[1] Bei Yan,[1] Rui Ge,[1] Jianlan Xie,[1] Yuchen Peng,[1] Hang Li,[1] Wei E. I. Sha,[2] and Jianjun Liu[1, *]

[1] *Key Laboratory for Micro/Nano Optoelectronic Devices of Ministry of Education & Hunan Provincial Key Laboratory of Low-Dimensional Structural Physics and Devices, School of Physics and Electronics, Hunan University, Changsha 410082, China*

[2] *Key Laboratory of Micro-nano Electronic Devices and Smart Systems of Zhejiang Province, College of Information Science and Electronic Engineering, Zhejiang University, Hangzhou 310027, China*

*Corresponding author: jianjun.liu@hnu.edu.cn

**Abstract:** Topological corner state (TCS) and topological edge state (TES) have provided new approaches to control the propagation of light. The construction of topological coupled cavity-waveguide system (TCCWS) based on TCS and TES is worth looking forward to, due to its research prospects in realizing high-performance micro-nano integrated photonic devices. In this Letter, TCCWS is proposed in two-dimensional (2D) photonic crystal (PC), which possesses strong optical localization, high quality factor and excellent robustness compared with the conventional coupled cavity-waveguide system (CCCWS). This work will provide the possibility to design high-performance logic gates, lasers, filters and other micro-nano integrated photonics devices and expand their applications.

## 1. Introduction

Controlling the propagation of light is the core content of the field of optical communication. With its unique band gap and local characteristics, photonic crystal (PC) has become an effective carrier for realizing on-chip integrated optical communication [1,2]. Coupled cavity-waveguide system (CCWS) based on PCs can realize the input, localization and output of optical signals, and has been widely used in filters [3,4], sensors [5,6], lasers [7,8] and other micro-nano integrated photonics devices. However, the performance of conventional optical waveguide and optical cavity is easily affected by obstacles and defects caused by actual fabrication errors.

Photonic topological insulators and their topological states have provided new materials and new approaches to control the propagation of light. Topological edge state (TES) [9-28] and topological corner state (TCS) [29-36] have been generated based on topological phases transition. The optical waveguide based on TES (i.e., TES waveguide), without backscattering and immune fabrication errors [9-12], has been applied to splitters [17], logic gates [18,21] and lasers [24,25], which significantly improved device performance. The optical cavity based on TCS (i.e., TCS cavity) can achieve strong localization without introducing defect structures [30], but its application is very limited by its inability to transmit optical signal. If TCS cavity and TES waveguide are combined, a unique CCWS named topological CCWS (TCCWS) will be formed, which will be topologically protected. The construction of TCCWS is worth looking forward to, since it possesses both strong localization and robust transmission, and possesses the possibility to design high-performance logic gates, lasers, filters and other micro-nano integrated photonic devices and expand their applications. Different from the previous CCWSs [18,24], which localized the energy within an annular region, this TCCWS will localize the optical signal at a point, so as to achieve stronger localization and smaller mode volume, which will be more conducive to the design of on-chip integrated optical communication devices.

In this Letter, the TES waveguide and the TCS cavity are proposed based on the bulk-edge-corner correspondence of the square lattice PC. Owing to the flat band characteristics of the edge of TES band, the TCCWS constructed with TCS cavity and TES waveguide has strong cavity-waveguide interaction. In contrast to the conventional CCWS (CCCWS) constructed with point-defect cavity and line-defect waveguide, the strong optical localization, high quality factor and excellent robustness of TCCWS are verified by theoretical investigation and numerical simulation.

## 2. Model and theory

The lattice constant of the square lattice PC is $a = 0.5$ μm, the initial side length of the square dielectric rod is $l_0 = 0.44a$, the refractive index $n_a = 3.47$, and the background refractive index $n_b = 1$. As shown in Fig. 1(a), the square lattice PC protected by inversion symmetry has four inversion centers, denoted as A, B, C, and D, respectively. As shown in Figs. 1(b) and 1(c), the four inversion centers correspond to four types of unit cell (UC) respectively. After the periodic array, four types of square lattice PCs can be obtained (only the shapes of scatterers at four boundaries of each PC are not the same). It can be seen from Fig. 1(d) that, although the arrangement of the scatterers in the four UCs is different, their band structures are the same and have a common band gap. By analyzing the symmetry of the electric field $E_z$ at the Γ and X (Y) points in the Brillouin zone, the corresponding Zak phases can be obtained, as shown in Fig. 1(e).

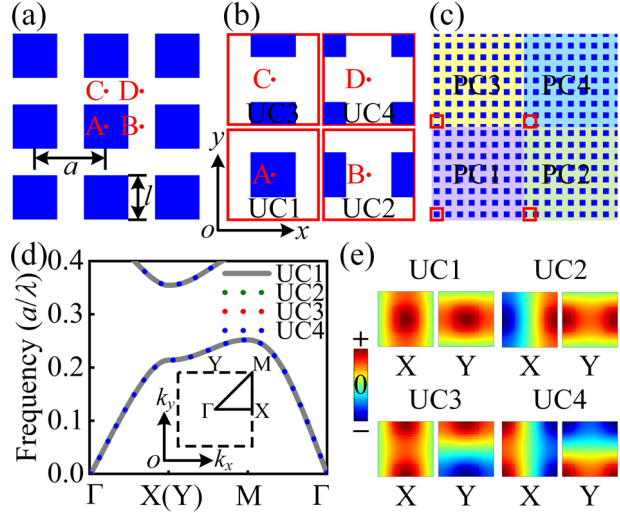

**Fig. 1.** (a) Square lattice PC. (b) Four types of UC. (c) Four kinds of square lattice PC constructed with periodic arrays of four UCs (red box). (d) The band structures of four UCs under TM polarization, among which the three dotted lines of green, red, and blue overlap, and only the blue dots are shown. (e) The electric field $E_z$ of four UCs at the high symmetry points of the first bulk band.

According to the bulk-edge correspondence, when the Zak phase of the PC bulk bands changes, there will be TES in the corresponding band gap [37]. When considering the first bulk band, the Zak phase of the two-dimensional (2D) PC along the $j$ direction is [13,31],

$$\theta_j^{Zak} = \int dk_x dk_y \text{Tr}[\mathbf{A}_j(k_x, k_y)], \quad j = x, y \tag{1}$$

where the integral range is the first Brillouin zone, and $\mathbf{A}_j(k_x, k_y) = i\langle \psi | \partial_{k_j} | \psi \rangle$ is the Berry connection. There is a corresponding relationship between the Zak phase and the 2D polarization (**P**) along the $j$ direction: $\theta_j^{Zak} = 2\pi P_j$ [13]. In the structure protected by inversion symmetry, the value of Zak phase is quantized and can only be 0 or π. In this Letter, the Zak phase is determined by the electric field $E_z$ of the eigenvalues at the high symmetry points in the Brillouin zone. Considering that the electric field $E_z$ at Γ point is mirror symmetric, if the electric field $E_z$ at X (Y) point along the $x$ ($y$) direction is mirror symmetric/antisymmetric, the Zak phase in this direction is 0 (trivial phase)/π (topological non-trivial phase) [12,31]. It can be seen from Fig. 1(e) that the 2D Zak phases of the four PCs are (0, 0), (π, 0), (0, π) and (π, π) respectively, and the corresponding 2D **P** is (0, 0), (1/2, 0), (0, 1/2) and (1/2, 1/2). The Zak phases of the four PCs are different. The reason is that the value of the Zak phase is related to the inversion center (the center of the UC), and the Zak phase difference of the UCs corresponding to the two inversion centers along the $j$ direction is π [37]. For two PCs with a Zak phase difference along the $x$ ($y$) direction, their combined structure will lead to TES at the interface along the $y$ ($x$) direction.

In order to observe the TES, PC1 is combined with PC3 and PC4 along the $x$ direction respectively, and the projected band structures are calculated as shown in Figs. 2(a) and 2(b).

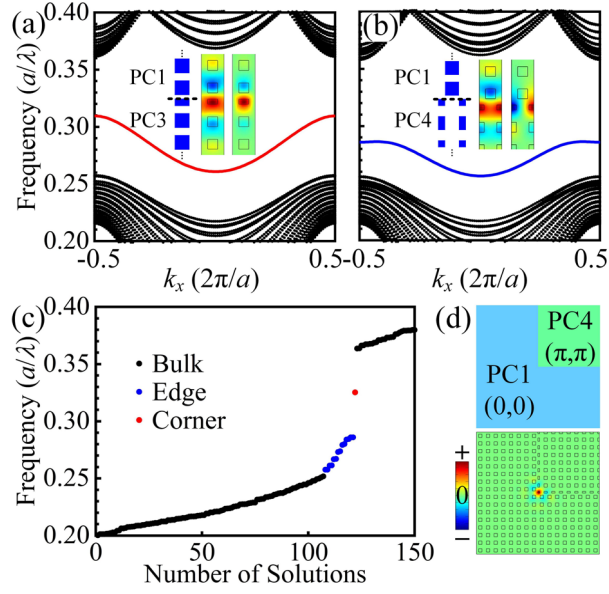

**Fig. 2.** (a)-(b) The projected band structures of PC1 combined with PC3 and PC4, respectively. The insets show the supercells and electric field $E_z$ of two TESs at the high symmetry points (0,0) (left) and ($\pi/a$,0) (right). (c) The eigenmodes of the combined structure of PC1 and PC4. (d) The schematic of the combined structure of PC1 and PC4 and the electric field $E_z$ of the TCS.

It can be seen from Figs. 2(a) and 2(b) that there are TESs (TES1 and TES2, respectively) in the band gap of the two PC combination structures. However, the topology of the TESs in Figs. 2(a) and 2(b) is not the same, which is related to edge polarization ( $p_j^{\text{edge}}$ ). Similar to bulk polarization, edge polarization can be determined by the symmetry of the electric field $E_z$ at high symmetry points in the Brillouin zone. From the illustrations in Figs. 2(a) and 2(b), at (0, 0), the electric field $E_z$ of the two TESs are mirror symmetric, and at ($\pi/a$, 0), TES1 is mirror symmetric, while TES2 is mirror antisymmetric. Therefore, the two edge polarizations along the $x$ direction are $p_{x(\text{TES1})}^{\text{edge}} = 0$ and $p_{x(\text{TES2})}^{\text{edge}} = 1/2$, respectively. If PC1 is combined with PC3 and PC4 respectively along the $y$ direction, the former will no longer have TES (i.e., $p_{y(\text{TES1})}^{\text{edge}} = 0$), while the latter will still have TES with $p_{y(\text{TES2})}^{\text{edge}} = 1/2$. The existence of the TCS depends on the edge polarization [38],

$$Q^{\text{corner}} = p_x^{\text{edge}} + p_y^{\text{edge}} \qquad (2)$$

According to Eq. (2), when $p_j^{\text{edge}} = 1/2$, $Q^{\text{corner}} = 1$, which corresponds to the existence of TCS, while $p_j^{\text{edge}} = 0$ cannot lead to TCS. By calculating the eigenmodes of the combined structure of PC1 and PC4, the existence of TCS can be verified, as shown in Figs. 2(c) and 2(d).

It can be seen from Fig. 2(c) that there has TCS in the band gap of the TES, and the corresponding normalized frequency is 0.325. Figure 2(d) shows the electric field $E_z$ of TCS. The electromagnetic energy is strongly located in the 90° corner formed by the combination of two PCs.

## 3. Results and discussion

In this Letter, TCCWS is constructed with the TCS cavity and TES waveguide. At the same time, the CCCWS is constructed with the point-defect cavity and line-defect waveguide. The performances of the two systems are explored and compared, as shown in Fig. 3.

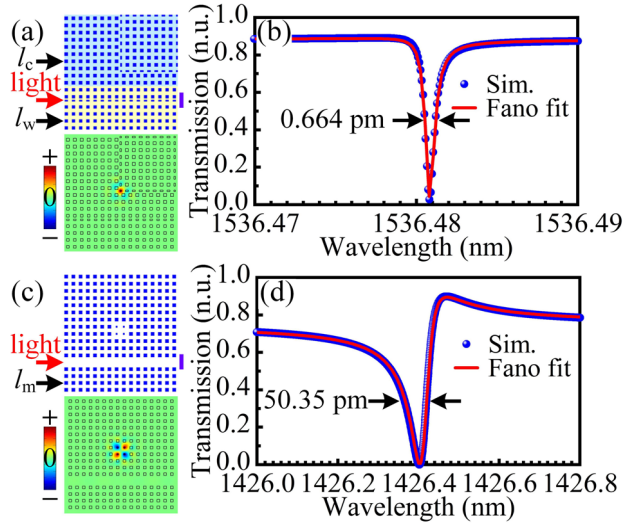

**Fig. 3.** (a) The schematic of TCCWS (top), the electric field $E_z$ of the resonance mode (bottom). The red arrow and the purple rectangle correspond to the light source and the monitor, respectively. (b) Transmission spectrum: simulated data points and Fano fitting. (c) The schematic of CCCWS (top), the electric field $E_z$ of the resonance mode (bottom). (d) Transmission spectrum: simulated data points and Fano fitting.

As shown in Fig. 3(a), the blue part is the TCS cavity, which is constructed with

PC1 and PC4, and the side length of the square dielectric rod is $l_c$ = 0.44$a$. The yellow part is the TES waveguide, which is constructed with PC1 and PC3, and the side length of the square dielectric rod is $l_w$ = 0.386$a$. The side length of the square dielectric rod of the CCCWS is $l_m$ = 0.4$a$, as shown in Fig. 3(c). The CCWS constructed with localization mode cavities and transmission mode waveguides will lead to Fano resonance with the line shape of sharp and asymmetry. The line shape of the Fano resonance can be expressed as [39-41],

$$F(\omega) = A_0 + F_0 \frac{[q + 2(\omega - \omega_0)/\Delta\omega]^2}{1 + [2(\omega - \omega_0)/\Delta\omega]^2} \quad (3)$$

where $A_0$ and $F_0$ are constants, $\omega_0$ is resonance frequency, $\Delta\omega$ is resonance linewidth, $q$ is Fano parameter, indicating the degree of asymmetry. It can be seen from Figs. 3(b) and 3(d) that the transmission spectrum data points calculated by simulation fit well with Fano line shape, and the resonance linewidth of TCCWS is much smaller than that of CCCWS, which corresponds to stronger localization.

The quality factor ($Q$) is an important performance parameter of CCWS, which can be expressed by the resonance frequency ($\omega_0$) and the decay time ($\tau$) of the electromagnetic energy in the cavity or the resonance linewidth ($\Delta\omega$) [42] as,

$$Q = \frac{\omega_0 \tau}{2} = \frac{\omega_0}{\Delta\omega} \quad (4)$$

$Q$ = 2.3139×10$^6$ (2.8329×10$^4$) of TCCWS (CCCWS) can be calculated from Eq. (4).

Next, further analyze the reason why the $Q$ of TCCWS is significantly better than that of CCCWS. The $Q$ in CCWS consists of two parts in time $\tau$, namely the radiation part $Q_r$ to the surrounding environment in time $\tau_r$ and the coupling part $Q_w$ to the waveguide in time $\tau_w$. The relationship between the parameters can be expressed as $1/Q = 1/Q_r + 1/Q_w$ and $1/\tau = 1/\tau_r + 1/\tau_w$ [42], where $\tau_r$ is determined by the material and structure of the cavity and can be qualitatively expressed as $\tau_r = f(a, \varepsilon, \delta, N)$, $a$ is the lattice constant, $\varepsilon$ is the dielectric constant, $\delta$ is the filling rate, and $N$ is the number of lattice periods around the cavity. The $\tau_w$ is determined by the material and structure of the waveguide and the distance ($d_{cw}$) between the cavity and the waveguide.

The material and structure of the waveguide determine the dispersion relationship and thereby the density of states DOS($\omega$) of the transmission mode. When the frequency of the transmission mode is equal to the resonance frequency of the cavity, that is, $\omega = \omega_0$, the waveguide mode and the cavity mode are coupled. Considering the DOS($\omega_0$) at the resonance frequency, its definition is [43,44],

$$\text{DOS}(\omega_0) = \left(\frac{1}{\pi}\right)\left(\frac{\partial k}{\partial \omega}\right)_{\omega=\omega_0} \tag{5}$$

where $\partial k/\partial \omega$ is the reciprocal of the slope of the dispersion curve (that is, the reciprocal of the group velocity). DOS($\omega_0$) represents the number of eigenmodes corresponding to the cavity resonance frequency, and the decay time is proportional to it, $\tau_w \propto \text{DOS}(\omega_0)$ [45]. The decay time is proportional to the decay distance, namely $\tau_w \propto d_{cw}$. Therefore, $\tau_w$ can be qualitatively expressed as, $\tau_w = f(\text{DOS}(\omega_0), d_{cw})$.

The values of the parameters $a$, $\varepsilon$, and $N$ in the TCS cavity proposed in this Letter are respectively equal to that of the conventional point-defect cavity, and $\delta$ is approximately equal, thereby the $\tau_r$ and $Q_r$ of the two systems are approximately equal. Considering $Q_r \gg Q_w$ in CCWS [42], it can be deduced from $1/Q = 1/Q_r + 1/Q_w$ that $Q$ mainly depends on $Q_w$.

Under the condition of $d_{cw} = 4a$, DOS($\omega_0$) of TCCWS and CCCWS is compared and discussed, as shown in Fig. 4.

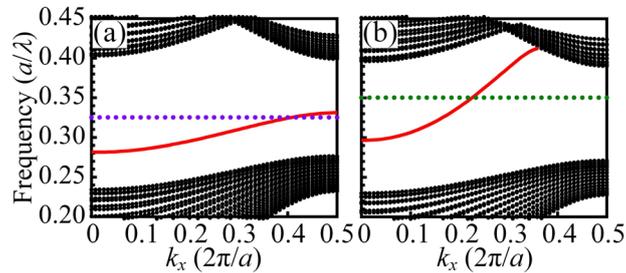

**Fig. 4.** (a) The projected band structure of the TES waveguide, the purple dotted line corresponds to the frequency of TCS. (b) The projected band structure of the line-defect waveguide, the green dotted line corresponds to the resonance frequency of point-defect cavity.

As shown in Figs. 4(a) and 4(b), the group velocity corresponding to the resonance frequency of TCCWS is smaller, that is, the value of $\partial k/\partial \omega$ is larger than that of CCCWS. According to Eq. (5), DOS($\omega_0$) is higher, and thereby $\tau_w$ and $Q_w$ are higher, thus the $Q$ of TCCWS is significantly better than that of CCCWS.

In addition to strong localization and high $Q$, TCCWS can also be immune to obstacles and defects caused by actual fabrication errors. The robustness ($R$) (anti-interference performance) of TCCWS and CCCWS is further explored. Under certain external disturbances, $R$ denotes the change of CCWS performance parameters, which can be qualitatively expressed as,

$$R_{\text{CCWS}} = f(\alpha, \beta, s) \tag{6}$$

where $\alpha = \Delta\lambda_0 / \lambda_0$, $\lambda_0$ is the resonance wavelength, and $\Delta\lambda_0$ is its offset; $\beta = \Delta Q / Q$, $Q$ is the quality factor of the system, and $\Delta Q$ is its offset; $s$ is the interference factor, corresponding to obstacle, defect, etc.

In this Letter, the $R_{\text{CCWS}}$ of TCCWS and CCCWS are compared in the case of introducing metal obstacle ($s_1$) in the waveguide and reducing the size of square dielectric rod introducing defect ($s_2$) in the cavity, as shown in Figs. 5(a)-5(d) and Table 1.

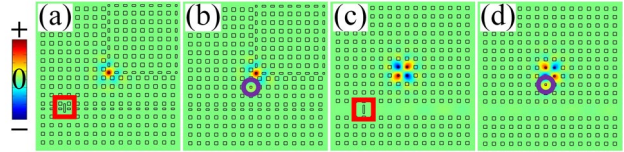

**Fig. 5.** (a)-(b) TCCWS: the resonance wavelengths are 1536.48 nm and 1534.71 nm, respectively and corresponding $Q$s are $2.2463\times10^6$ and $1.4117\times10^6$. (c)-(d) CCCWS: the resonance wavelengths are 1426.40 nm and 1415.77 nm, respectively and corresponding $Q$s are $1.2952\times10^4$ and $1.0546\times10^4$. The red rectangle corresponds to the metal obstacle ($s_1$), and the purple circle corresponds to the defect ($s_2$).

Table 1. Comparison of robustness between TCCWS and CCCWS

| CCWS | $α(s_1)$ | $β(s_1)$ | $α(s_2)$ | $β(s_2)$ |
|---|---|---|---|---|
| TCCWS | 0 | 2.9% | 0.1% | 39.0% |
| CCCWS | 0 | 54.3% | 0.7% | 62.8% |

It can be seen from Figs. 5(a)-5(d) and Table 1 that $α(s_2)$, $β(s_1)$ and $β(s_2)$ of TCCWS (CCCWS) have small (large) changes under the same interference conditions. Although TCCWS is also affected by fabrication errors, its affected degree is far lower than that of CCCWS, which reflects a relatively excellent anti-interference performance. Therefore, the TCCWS proposed in this paper possesses excellent applicability for micro-nano integrated photonic devices that require strong localization, high $Q$ (i.e., narrow linewidth) and stable resonance frequency, such as filters [3,4], lasers [7,8,24], etc.

## 4. Conclusion

In this Letter, TCCWS is constructed with TCS cavity and TES waveguide, and its resonance characteristics and DOS are investigated. In contrast to the CCCWS, it is proved that the TCCWS possesses strong optical localization, high $Q$ and excellent robustness, which provides a new approach to control the propagation of light and a new platform to realize high-performance micro-nano integrated photonic devices.

## Acknowledgements

This work was supported by the National Natural Science Foundation of China (Grant Nos. 61405058 and 62075059), the Natural Science Foundation of Hunan Province (Grant Nos. 2017JJ2048 and 2020JJ4161), and the Fundamental Research Funds for the Central Universities (Grant No. 531118040112). The authors acknowledge Professor Jian-Qiang Liu for software sponsorship.